\pdfoutput=1

\documentclass{jpsj-suppl}
\usepackage[footskip=0.35in,bottom=0.75in]{geometry}
\usepackage{txfonts} 
\usepackage{physics}
\usepackage{tikz,pgfplots,textcomp}
\usepackage{graphicx}
\usepackage[utf8]{inputenc}

\title{Astrophysical Neutrinos and Flavor Evolution With Self Interaction}

\author{Taygun \textsc{Bulmus}$^{1}$ and Yamac \textsc{Pehlivan}$^{1}$}

\inst{$^{1}$Department of Physics, Mimar Sinan Fine Arts University, Sisli, Istanbul, 34380, Turkey}

\email{ismail.taygun.bulmus@ogr.msgsu.edu.tr}

\abst{In a core collapse supernova, collective oscillations of neutrinos emitted by
the proto-neutron star significantly modifies the partition of energy
between different flavors in a way which is potentially important for the
nucleosynthesis of heavy elements at the outer layers. As the outcome of
collective oscillations 
depends nonlinearly on the chosen initial conditions proto-neutron star, their 
systematical study across the proto-neutron star
model space may provide answers to some key questions.}

\kword{supernova neutrinos, collective neutrino oscillations, spectral split}

\begin{document}
\maketitle

Core collapse supernovae are the most extreme neutrino events in the Universe.
The gravitational binding energy of the collapsed matter is emitted almost
entirely in terms of neutrinos by the newly formed proto-neutron at the center
as it rapidly cools down in the next few seconds
\cite{Colgate:1966ax,Woosley:1986ta,1989ARA&A..27..629A,Kotake2005}.  These
neutrinos are thought to play important roles in the dynamics of the exploding
star \cite{Takiwaki:2013cqa} , as well as in the $r$-process \cite{Woosley:1994ux}
and $\nu$-process \cite{Epstein:1988gt,Woosley:1989bd} nucleosynthesis at the
outer layers.  Neutrinos freely stream inside the supernova once they decouple
from the proto-neutron star, but the energy partition between different neutrino
and anti-neutrino flavors change as neutrinos undergo flavor evolution during
their propagation. The flavor evolution is strongly influenced by the neutrino
scattering in the matter despite the small cross sections at the MeV energies
relevant for the supernova because tiny neutrino scattering amplitudes
coherently superpose to create a macroscopic refractive effect on neutrino
propagation \cite{Wolfenstein:1977ue,Mikheev:1986wj}. At the high neutrino
densities near the proto-neutron star, neutrino-neutrino scattering also
contributes to the refraction \cite{fuller&mayle,savage&malaney} in which case
identical particle effects couple the flavor evolutions the interacting
neutrinos \cite{Pantaleone:1992xh,Pantaleone:1992eq}, turning the neutrino
flavor evolution into a many-body problem
\cite{Sigl:1992fn,Duan:2010bg,Bell:2003mg,Friedland:2003eh,Friedland:2003dv,Balantekin:2006tg,Pehlivan:2011hp,Vaananen:2013qja,Volpe:2013jgr}.
Several emergent phenomena resulting from the many-body nature of this
\emph{neutrino self refraction} were identified such as synchronized
oscillations, \cite{Samuel:1993uw, Kostelecky:1994dt}, bi-polar oscillations
\cite{Samuel:1995ri}, and spectral splits (or swaps) \cite{Duan:2006an} which are now
generally referred to as \emph{collective neutrino oscillations}.  

\begin{figure}
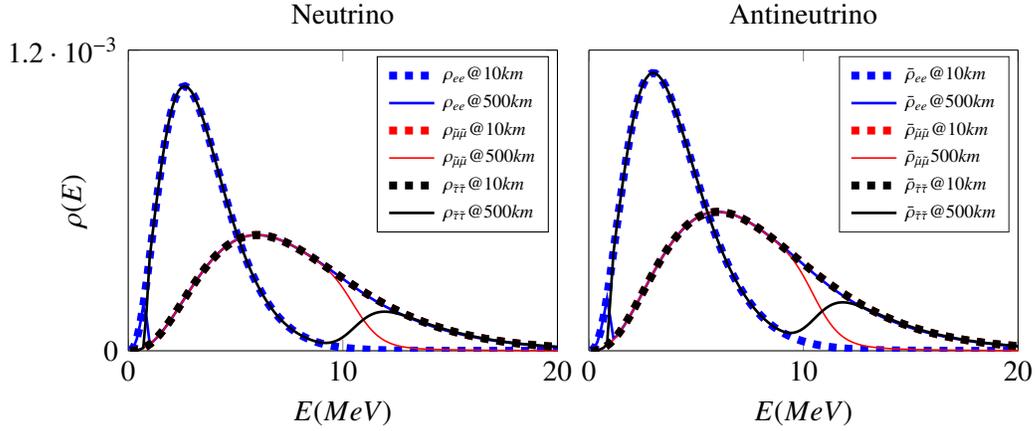

\begin{minipage}{0.45\textwidth}
		\newlength\figureheight 
		\newlength\figurewidth 
		\setlength\figureheight{4cm} 
		\setlength\figurewidth{6cm}
		\input{three1.tikz}
\end{minipage}
\begin{minipage}{0.45\textwidth}
		\setlength\figureheight{4cm} 
		\setlength\figurewidth{6cm}
		\input{antithree1.tikz}
\end{minipage}
\caption{Initial conditions with $T_{\nu_{e}}=3.5$ MeV,
$T_{\bar{\nu}_{e}}=4.0 $ MeV, $T_{\nu_x}=T_{\bar{\nu}_x}=8.0$ MeV. We take 
number fluxes to be the same for all neutrino and antineutrino flavors. 
Dashed lines show the initial neutrino spectra at the proto-neutron star with
a $10$ km radius. Continuous lines show the final neutrino spectra at $500$ km
away from the center.}
\label{Full_Graph}
\end{figure}
\begin{figure}
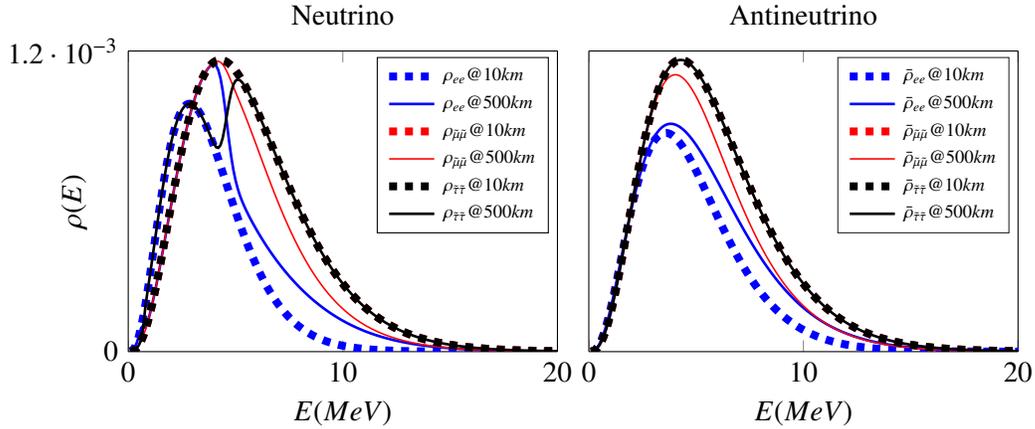

\begin{minipage}{0.45\textwidth}
		\setlength\figureheight{4cm} 
		\setlength\figurewidth{6cm}
		\input{partial2.tikz}
\end{minipage}
\begin{minipage}{0.45\textwidth}
		\setlength\figureheight{4cm} 
		\setlength\figurewidth{6cm}
		\input{antipartial2.tikz}
\end{minipage}
\caption{Initial conditions with $ T_{\nu_{e}}=3.8 $ MeV, $T_{\bar{\nu}_{e}}=4.8$ MeV, 
$T_{\nu_x}=T_{\bar{\nu}_x}=5.7$ MeV.  
neutrino number fluxes are taken to be unequal with slightly more muon and tau flavor
(anti) neutrinos than electron (anti) neutrinos.  
Dashed lines show the initial neutrino spectra at the proto-neutron star with
a $10$ km radius. Continuous lines show the final neutrino spectra at $500$ km
away from the center.
} 
\label{Partial_Graph}
\end{figure}

As a nonlinear many-body system, flavor evolution of neutrinos near the
proto-neutron star seem to be sensitive to various factors. For example, the
influence of the geometry is investigated by various groups
\cite{Fogli2007,Abbar:2015mca,Raffelt:2013rqa}. Our aim is to systematically
investigate the dependence of the collective neutrino oscillations on the
initial conditions, i.e., the energy spectra of all neutrino and anti neutrino
flavors at the point where they thermally decouple from the proto-neutron star
and begin freely streaming inside the supernova. Since different neutrino
flavors undergo different interactions inside the proto-neutron star, they
thermally decouple from different radii with unequal temperatures. Although
model independent arguments give us a temperature hierarchy of $T_{\nu_e}<T_{\bar{\nu}_e}<T_{\nu_x}=T_{\bar{\nu}_x}$ where $x$ denotes both
muon and tau flavors, the actual values of these temperatures are significantly
model dependent. For a recent compilation of different results from neutrino
transport simulations inside the proto-neutron star, see Ref.
\cite{Mathews2014}. Number fluxes for different flavors can also differ slightly
from one another \cite{Buras:2002wt,Dasgupta:2010cd}.

Figs. \ref{Full_Graph} and \ref{Partial_Graph} show the results of collective
neutrino oscillations for two different sets of initial conditions on the
proto-neutron star. Both figures are obtained by treating the self refraction of
neutrinos under the mean field approximation \cite{Sigl:1992fn} and by adopting
the neutrino bulb model which is the simplest effective treatment of the
geometry of intersecting neutrino trajectories \cite{Duan:2006an}. The
background density of other particles is assumed to be constant and their
refractive effect is treated by using effective mixing parameters in matter
\cite{Kuo:1989qe,Agashe:2014kda}.  We assume inverted mass hierarchy. Note that
since $\nu_\mu$ and $\nu_\tau$ are emitted with almost the same initial spectrum
from the proto-neutron star, one can conveniently work in a rotated flavor basis
$\tilde{\nu}_\mu$ and $\tilde{\nu}_\tau$, which mix with $\nu_e$ but not with
each other (see, e.g., Ref. \cite{Pehlivan:2014zua}). 

The results shown in Figs. \ref{Full_Graph} and \ref{Partial_Graph} are
qualitatively different from each other: in Fig. \ref{Full_Graph}, there is a
complete spectral swap at high energies, i.e., the energy spectra are exchanged
between the flavors after the collective oscillations. In Fig.
\ref{Partial_Graph} the swap is only partial. It is clear that even in this
simplest scenario, we do not fully understand how the behaviour of self
interacting neutrinos, as a nonlinear many-body system, depend on the initial
conditions.  We believe that it may be useful to carry out a systematical study
of the proto-neutron star model space and identify the response of the neutrino
collective oscillations in different regions.  In studies of $r$ or
$\nu$-process nucleosynthesis in the supernova, one usually adopts certain
initial neutrino spectra and calculates the isotopic yields. It is interesting
to consider the reverse problem and ask if there is a set of initial conditions
favorable for the nucleosynthesis by scaning the model space. Recently a formal
correspondence is established between the flavor evolution of self interacting 
neutrinos and the dynamics of fermionic Cooper pairs
\cite{Pehlivan:2011hp,Pehlivan:2014zua,Pehlivan:2016lxx} leading to the
identification of some dynamical symmetries. Since dynamical symmetries and the
associated invariants restrict the motion of the system in the phase
space, they can possibly be used to relate the behaviour of neutrinos to the initial
conditions on the proto-neutron star. 

\appendix{This work was supported by the Scientific and Technological  Research  Council  of  Turkey under project number 115F214.}
    	\bibliographystyle{jpsj}

\end{document}